# Information Retrieval on the Web and its Evaluation

Deepak Garg
Associate Professor
CSED, Thapar University
Patiala, India

Deepika Sharma
Student- M.E.
CSED, Thapar University
Patiala, India

## ABSTRACT
Internet is one of the main sources of information for millions of people. One can find information related to practically all matters on internet. Moreover if we want to retrieve information about some particular topic we may find thousands of Web Pages related to that topic. But our main concern is to find relevant Web Pages from among that collection. So in this paper I have discussed that how information is retrieved from the web and the efforts required for retrieving this information in terms of system and users efforts.

## 1. INTRODUCTION
The Web has undergone an exponential growth in the past few years. It has been estimated that there are approximately 15-20 billion pages present on the Web and recently this count has hit the mark of 1 trillion. According to the studies only 80-85 % of the total Web pages that are available on the Web give useful information and the remaining 20-15% are mostly duplicates of the original pages or near duplicates and some of them are completely irrelevant pages. Thus, the Web explosion offers lots of new problems for the information retrieval systems. These information retrieval systems help users complete the search tasks, by finding a handful of relevant documents among thousands and thousands of pages of text with little structural organization. At the same time, developers of retrieval systems must be able to evaluate the overall effectiveness of these systems i.e., the relevance of results it retrieves in response to a user query.

## 2. INFORMATION RETRIEVAL ON THE WEB
Information Retrieval on the Web has always been different and difficult task as compared with a classical information retrieval system (Library System). To explain the difference between classical information retrieval and information retrieval on the Web we compare the two. Basically the differences can be partitioned into two parts, namely differences in the documents and differences in the users.

We first discuss the differences in the documents.

- **Hypertext**: Documents present on the web are different from general text-only documents because of the presence of hyperlinks. It is estimated that there are roughly 10 hyperlinks present per document.
- **Heterogeneity of document**: The contents present on a web page are heterogeneous in nature i.e., in addition to text they might contain other multimedia contents like audio, video and images.
- **Duplication**: On the Web, over 20% of the documents present are either near or exact duplicates of other documents and this estimation has not included the semantic duplicates yet.
- **Number of documents**: The size of Web has grown exponentially over the past few years. The collection of documents is over trillions and this collection is much larger than any collection of documents processed by an information retrieval system. According to estimation, Web currently grows by 10% per month.
- **Lack of stability**: Web pages lack stability in the sense that the contents of Web pages are modified frequently. Moreover any person using internet can create a Web pages even if it contains authentic information or not.

The users on the Web behave differently than the users of the classical information retrieval systems. The users of the latter are mostly trained librarians whereas the range of Web users varies from a layman to a technically sound person. Typical user behaviour shows:

- **Poor queries**: Most of the queries submitted by users are usually short and lack useful keywords that may help in the retrieval of relevant information.
- **Reaction to results**: Usually users don't evaluate all the result screens, they restrict to only results displayed in the first result screen.
- **Heterogeneity of users**: There is a wide variance in education and Web experience between Web users.

Thus, the main challenge of information retrieval on the Web is how to meet the user needs given the heterogeneity of the Web pages and the poorly made queries.

## 3. IR (INFORMATION RETRIEVAL) TOOLS ON THE WEB
Information from Web can be retrieved by number of tools available ranging from General Purpose Search Engines to Specialized Search Engines. Following are the most commonly used Web IR tools:

- **General-Purpose Search Engine**: They are the most commonly used tool for information retrieval. Google, AltaVista, Excite are some of the examples.

Each of them has its own set of Web pages which they search to answer a query.
- **Hierarchical directories**: In this approach the user is required to choose one of a given set of categories at each level to get to the next level. For example, Yahoo! or the dmoz open directory project
- **Specialized Search Engines**: These Search Engines are specialized on an area and provides huge collection of documents related to that specific area. For e.g. PubMed, a Search Engine specialized on medical publications. It offers collection of millions of research papers, articles; journals related to bio medical sciences, life sciences etc [1].

## 4. GENERAL PURPOSE SEARCH ENGINE

General Purpose Search Engines are used to index a sizeable portion of the Web across all topics and domains to retrieve the information. Each such Engine consists of three major components:
- A spider or crawler [5] browses the Web by starting with a list of URLs called the seeds. As the crawler visits these URLs, it identifies all the hyperlinks in the page and adds them to the list of URLs which are visited recursively to form a huge collection of documents called corpus. The corpus is typically augmented with pages obtained from direct submissions to search engines and various other sources. Each crawler has different policies with respect to which links are followed, how deep various sites are explored, etc. As a result, there is surprisingly little correlation among corpora of various engines [8].
- The indexer processes the data and represents it usually in the form of fully inverted files. However, each major Search Engine uses different representation schemes and has different policies with respect to which words are indexed.
- The query processor which processes the input query and returns matching answers, in an order determined by a ranking algorithm. It consists of a front end that transforms the input and brings it to a standard format and a back end that finds the matching documents and ranks them.

## 4.1 A Brief History of Search Engines
Search Engines have evolved a lot since their inception. This evolution witnessed three major generations; each generation considered its own approach for retrieving of relevant documents. Following are the three main generations:

- **1$^{st}$ Generation:** This generation came around 1996. It search ranked sites based on page content. Documents are treated as collection of words and no importance is given on semantics of the documents. The main disadvantage of this generation was that any document can be made relevant by keyword stuffing so as to increase the content similarity- examples are Excite, Alta Vista, and Infoseek.
- **2$^{nd}$ Generation**: This generation relies on contents and as well as on link analysis for ranking- so they take the structure of the Web as a graph into account. It considers site popularity as the criteria for ranking the document as relevant. But this approach too has its flaws like spammers can create link farms i.e. heavily interconnected site which may make any document or page of lesser importance more important. For example Lycos.
- **3$^{rd}$ Generation**: Apart from page contents and web structure this generation considers page reputation as one of the major criteria. According to this approach if a page is referred by a highly reputed page then it has more relevance, more inlinks to a page means that the page has high reputation. Examples of 3$^{rd}$ generation search engines are Google, Yahoo!.

From the above discussion we inferred that the main task of a search engine is to retrieve information for a user query. To make this retrieval more relevant number of approaches is used as discussed above. But the best and universally accepted approach is to rank a page according to its relevance, this approach called as Page Rank is discussed below.

## 4.2 Ranking
Ranking is used to order the answer to a query in decreasing order of value. For this a numerical value called score is assigned to each document and the documents are arranged in the decreasing order of the score. This score is typically a combination of two criteria's query-independent and query-dependent criteria.
A query-independent criterion [1] assigns an intrinsic value to a document, regardless of the actual query by considering the publication data (like the site to which it belongs, the date of the last change, etc.), the number of citations (in degree), etc. A query-dependent criterion is a score which is determined only with respect to a particular query.

## 4.3 Graph Structure of Web
Before we study the details of each criterion we must represent Web as a directed graph [10], where each node represents a page and any link from one page to another page represents an edge i.e. if a page u contains a hyperlink for page v then that link is represented by a directed edge (u, v). Every page on the web has some number of forward links called as out edges and some number of back links called as in edges [15]. The number of out edges can be easily found by considering all the hyperlinks present at that page but it is difficult to find all the in edges to a page i.e. to find all the pages pointing to that page. For example in the figure 1 page B has two back links.

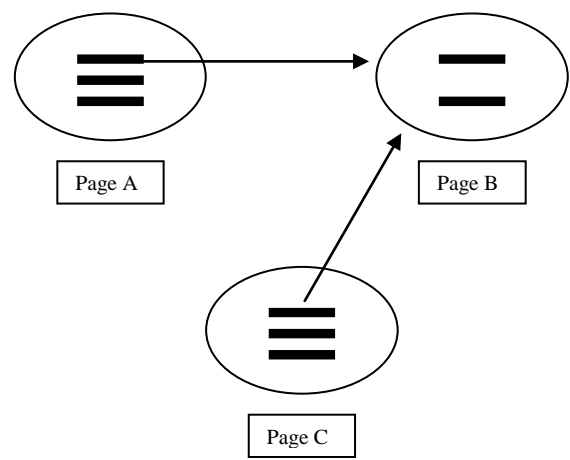

**Figure 1 A and C are the back links of B**

**Query-independent ranking criterion:** According to this criterion if a web page has larger number of hyperlinks pointing to it (also called inlinks) then it is considered as a better page.

The main drawback of this criterion is that each link is equally weighted. Thus, it cannot distinguish the quality of a page that gets pointed to by i low-quality pages from the quality of a page that gets pointed to by i high-quality pages. Obviously it is easier to make a page appear to be high-quality- just creates many other pages that point to it.

To remedy this problem, Brin and Page [2] invented the Page Rank measure. Page Rank is defined as follows:

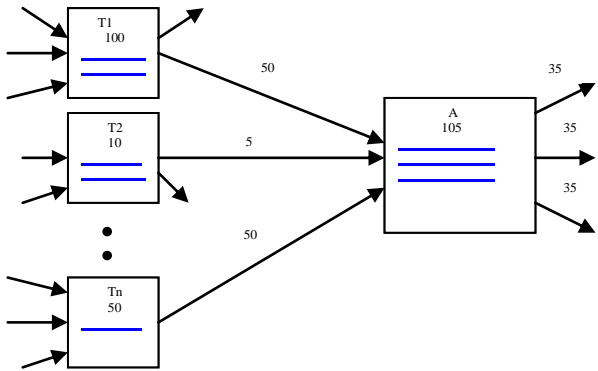

**Figure 2 Simplified Page Rank Calculation**

Consider that pages T1, T2….Tn are pointing to page A and C(T1) gives us the no of links going out of page T1 and so on then Page Rank of a page A is given as follows: We assume page A has pages T1...Ten which point to it (i.e., are citations). The parameter d is a damping factor which can be set between 0 and 1. We usually set d to 0.85. Also C (A) is defined as the number of links going out of page A. The Page Rank of a page A is given an iterative formula as:

PR(A) = (1-d) + d(PR(T1)/C(T1) +….+PR(Tn)/C(Tn))

Note that the Page Ranks form a probability distribution over web pages, so the sum of all web pages' Page Ranks will be one [15].

**Query-dependent ranking criterion:** It was developed by Kleinberg [1] [3]. It is described as follows:

For given a user query, the algorithm first constructs a graph specific for that query which is a sub graph of the main graph representing Web. In this query specific graph, nodes represent the pages and edges represent the hyperlink. For each page two types of scores are calculated: Authority Score and Hub Score. If a Web page has more relevant contents then its authority score is more and if a Web page contains hyperlinks to relevant pages then it has more hub score.

To begin the ranking of Web pages we consider $\forall p$, auth(p) = 1 and hub(p) = 1 where p represents the Web pages. We consider two types of updates: Authority Update Rule and Hub Update Rule. In order to calculate the hub/authority scores of each node, repeated iterations of the Authority Update Rule and the Hub Update Rule are applied. For a k-step application of the Hub-Authority algorithm apply first the Authority Update Rule and then the Hub Update Rule k times and then normalization is applied to finally converge the values of authority and hub score.

1) Authority Update Rule

$\forall p$, we update auth(p) as follows:

$$\sum_{i=1}^{n} hub(i)$$

According to Authority Update Rule if a page p is pointed by n number of pages, then authority score of that page is the sum of all the Hub scores of the pages that point to it

2) Hub Update Rule

$\forall p$, we update hub(p) as follows:

$$\sum_{i=1}^{n} auth(i)$$

Hub Update Rule states that if a page p contains hyperlinks for n number of pages then hub score of that page is the sum of the authority scores of all the pages to which it is linked.

3) Normalization

The final value of hub-authority scores of nodes is determined after infinite repetitions of the algorithm. Iteratively applying the Hub Update Rule and Authority Update Rule leads to diverging values. Thus the values obtained from this process will eventually converge. [3]

We summarize the above mentioned steps in the following algorithm:

1. Consider N be the number of nodes (pages) in the query specific graph.

2. For all n in the set N, let H[n] represents its hub score and A[n] represents its authority score.

3. Initialize the value of both H[n] and A[n] to 1, for all the nodes.

4. While the values of H[n] and A[n] does not converge perform the following steps:

   - For all n in N, A[n] = $\sum_{i=1}^{n} H(i)$

   - For all n in N, H[n] = $\sum_{i=1}^{n} A(i)$

   - For all n in N, normalize the value of H[n] and A[n]

## 4.4 Duplicate Filtering
Experiments indicate that over 20% of the publicly available documents on the Web are duplicates or near – duplicates [6]. There is a need to adopt some approach to find these duplicate documents, as discussed in [4] we can calculate the resemblance among Web pages in terms of a set intersection problem. The reduction to a set intersection problem is done via a process called shingling.

In this each document is viewed as a sequence of tokens. We can take tokens to be letters, or words, or lines. We assume that we have a parser program that takes an arbitrary document and reduces it to a canonical sequence of tokens. "Canonical" here means that any two documents that differ only in formatting or other information that we chose to ignore, for instance punctuation, formatting commands, capitalization, and so on, will be reduced to the same sequence.

A contiguous subsequence of w tokens contained in D is called a shingle. Given a document D, we can associate its w-shingling defined as the set of all shingles of size w contained in D. So for instance the 4-shingling of

(internet, scaled, data, storage, and, analysis)

is the set

{(internet, scaled, data, storage), (scaled, data, storage, and), (data, storage, and, analysis)}

Thus with each shingle a numerical score is associated which acts as a unique id for a particular shingle. This approach is called as fingerprinting. After fingerprinting each shingle in a document, the document gets an associated set of natural number as unique ids for all the shingles. For example, if D is a document the S(D) will contain set of all unique ids and size of S(D) is approximately equal to the number of words in the document D.

To calculate the resemblance between two documents A and B, we define r (A, B) as the resemblance factor and is calculated as below:

$$r(A,B) = \frac{|S(A) \cap S(B)|}{|S(A) \cup S(B)|}$$

Here, *r* = resemblance factor between two documents

$\cap$ = intersection operator

$\cup$ = union operator

Experiments seem to indicate that high resemblance (that is close to 1) captures well the informal notion of "near-duplicates" or "roughly the same".

## 5. QUANTIFYING THE QUALITY OF RESULT

The result that we get from any information retrieval system needs to be evaluated to see how relevant it is. Thus, there is a need to quantify the quality of result using some evaluation measures. This type of evaluation can be done by submitting a batch of pre-fabricated queries to the system and measure the relevance of results.

### 5.1 Related work

The original system-based evaluations were the Cranfield tests done in the 1950s and 1960s by Cyril Cleverdon, a librarian and computer scientist in the College of Aeronautics at Cranfield, UK. Cleverdon identified two broad types of "devices" that affect effectiveness in different ways; he called those that increased the proportion of relevant documents among those retrieved "precision devices" and those that increased the proportion of all relevant documents found "recall devices" [11]. Precision and recall devices could be combined in different ways to vary system behaviour in response to user queries; the challenge was measuring the effect of any given combination.

These tests done by Cleverdon were one of the first system evaluation tests, later many other organisations performed more evaluations like Text REtrieval Conference (TREC), organized by researchers at NIST since 1992, performs system-based evaluations [12], as do similar evaluation venues such as NTCIR (NII Test Collections for Information Retrieval, organized by the National Institute of Informatics in Japan), CLEF (the Cross-Language Evaluation Forum organized by the Istituto di Scienza e Tecnologie dell'Informazione), FIRE (the Forum for Information Retrieval Evaluation organized by the Information Retrieval Society of India), and INEX (the INitiative for the Evaluation of XML Retrieval).

### 5.2 Test Collection

Before starting the evaluation of an information retrieval system we need to understand that a user uses these systems for retrieval task like he may want to find all relevant documents for a query, to filter the relevant documents from the retrieved result set etc. All these retrieval tasks are done from a vast collection of documents called as test collection. A test collection encapsulates the experimental environment. It is meant to model users with information needs that are particular instances or examples of the task. These information needs are generally treated as if they do not change over time; if they are representative of the needs of users of the system in general, then showing that a system can perform well on them suggests that a system will perform well.

Test collections have three components:
- A corpus of documents to search;
- A set of user information needs;
- Judgement of the relevance of information needs to documents in the corpus.

### 5.3 Relevance Judgement

The relevance judgments tell us which documents are relevant to each of the information needs. As described above, since it is people that will be using the documents, relevance is something that must be determined by people. The system itself can only try to predict relevance; an evaluation determines how good the system is at predicting what will be relevant, and an experiment tells us whether one system is better at it than another. Once the topics have been finalized, human assessors can start judging documents for relevance. Assessors read documents, compare them to the topic definition, and say whether they are relevant or not (or possibly how relevant they are).

**Exhaustively judging** relevance—that is, judging every single document in the corpus to every single topic—is the only way to guarantee that all relevant documents are known. This is often impossible due to time and budget constraints, however. One assessor judging a million documents at a relatively quick rate of 10 per minute would take over ten months of 40-hour weeks to complete just one topic.

**Focusing judgment** effort on a small portion of the complete corpus can usually provide enough of the relevant documents for most evaluation and experimentation purposes. One simple approach is the pooling method: each topic in the collection is submitted to a variety of different retrieval systems, and the top *N* ranked documents from all of those systems are pooled for judging.

### 5.4 Evaluation Measures

Once a test collection has been finalized, at any time someone may submit a query derived from one of its topics to a retrieval system, obtain the ranked list of retrieved documents, and measure the system's effectiveness using the relevance judgments for that topic. The IR literature is awash with different evaluation measures meant to measure different aspects of retrieval performance; we will focus on a few of the most widely used.

### 5.4.1 Precision and Recall

Two of the most basic and most important aspects of effectiveness centre on the number of relevant documents retrieved:

1. Precision: The total number of relevant documents in the retrieved set gives us the precision of the system.

2. Recall: The total number of relevant documents retrieved from the total collection of documents or from the available corpus gives us the recall value for the system.

Suppose a system retrieves 10 documents from a corpus of one million; looking at our relevance judgments, we find that these 10 have been judged as follows: rel, rel, rel, rel, rel, rel, nonrel, nonrel, rel, rel. There are 162 known relevant documents in the corpus. The precision of these results is $8/10 = 0.8$ and the recall is $8/162 \approx 0.05$.

One solution is to look at precisions and recalls over a series of different rank cut offs. Rather than look at the entire retrieved set (which will likely be quite large, possibly the entire collection), we pick a rank cut off. Trends in precision and recall become apparent over a series of rank cut offs. In general, we define precision and recall at rank cut off *k* as

$$\text{precision@k} = \frac{\text{\# documents retrieved and relevant upto rank k}}{k}$$

$$\text{recall@k} = \frac{\text{\# documents retrieved and relevant upto rank k}}{\text{\#relevant documents}}$$

### 5.4.1.1 Precision – Recall Curve

Plotting recall and precision over a series of rank cut-offs produces the precision-recall curve. To understand the behaviour of precision-recall curve, we calculate the value of precision and recall at different ranks. For example consider the above mentioned case in which the system retrieves 10 documents. Suppose instead of 10 documents our system retrieves 50 documents out of which 20 are relevant, then precision =20/50 =0.4 and recall =20/162 $\approx$ 0.05. Here we see as the rank increases the value of precision decreases and value of recall increases this is because of the increase in number of retrieved documents. Using raw values of precision and recall at every possible rank cut-off produces a jagged curve like the one shown in Figure 3. This jagged curve represents that recall can never decrease with rank cut-off, while precision increases with every increase in recall and decreases while recall stays constant.

To produce a smoother curve we use a technique called *interpolation*. Interpolated precision is defined by a value of recall rather than by a rank cut-off; specifically, for a given recall level *r*, interpolated precision at *r* is defined to be the maximum measured precision at any rank cut-off *k* at which recall is no less than . We formulate this as

$$\text{i-precision@r} = \max_{k \text{ s.t. recall@k} \geq r} \text{precision@k}$$

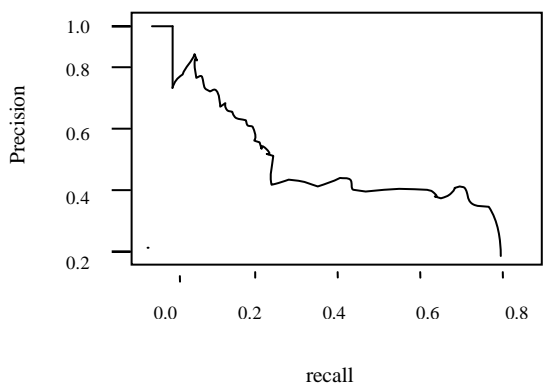

**Figure 3 Precision–Recall Curve**

There are 162 total relevant documents, so recall increases in increments of 1/162 = 0.006. Precision initially trends steadily downwards as recall increases from 0 to about 25, then holds steady as recall increases from 0.25 to about 0.7, after which it begins to fall again.

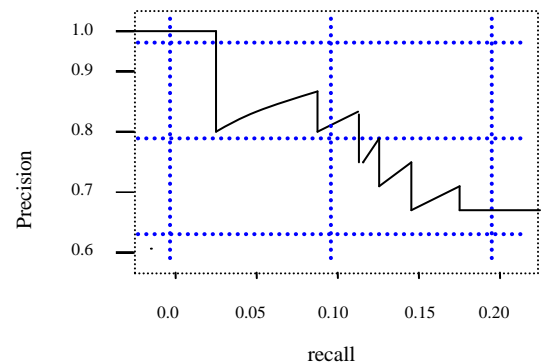

**Figure 4 Interpolating precision at recall points r = 0.0, 0.1, 0.2(details of Figure 3).**

First we locate point r on the x-axis (vertical dashed lines), then find the maximum value of precision after that point (horizontal dashed lines). That value is the interpolated precision at r, illustrated as solid lines. The details on Precision- Recall Curve can be read from [16].

### 5.4.2 Modelling User Effort

One factor of system performance that precision and recall-based measures do not directly address is the amount of effort a user can be expected to put in while interacting with the system. There are various families of measures that attempt to address this; the most commonly used are the discounted cumulative gain (DCG) family.

### 5.4.2.1 Discounted Cumulative Gain Family

Discounted cumulative gain (DCG) is defined by a gain function and a discount function. The gain function tells us the value of a particular relevant document to a user, allowing DCG to take advantage of grades of relevance. For instance, relevance judgments may be made on a three-point scale (not relevant, relevant, highly relevant) or a five-point scale (poor, fair, good, excellent, perfect); DCG's gain function can take advantage of these grades by mapping them to numeric values to reflect their utility to a user. Traditional precision and recall can only use binary judgments.

Two typical gain functions are the linear and exponential functions. Linear gain simply assigns incrementally increasing values to each relevance grade, e.g. nonrelevant→0, relevant→1, highly relevant →2. Exponential gain multiplicatively increases values, e.g. poor→0, fair→1, good→3, excellent→7, perfect→15.

By tuning the gain function, a developer can model users that have varying degrees of preference for different grades of relevance. The discount function reflects the patience a user has for proceeding down the ranked list. It is assumed that as the rank increases the gain function is likely to increase and discounts never increase or increase by a small margin.

Once a gain function *g* and a discount function *d* have been defined, we can define the discounted gain at any rank as the ratio of the gain of the document at that rank to the discount of that rank:

$$\text{Discounted Gain}@k = \frac{g(rel_k)}{d(k)}$$

DCG@k is then defined as the sum of the discounted gains from ranks 1 to *k*:

$$DCG@k = \sum_{i=1}^{k} \frac{g(rel_i)}{d(i)}$$

So we see that with the increase in rank value gain function behaves linearly and the discount function behaves logarithmically.

## 6. CONCLUSION
In this paper, we discussed Web information Retrieval methods and tools that take advantage of the Web particularities to mitigate some of the difficulties that Web information retrieval encounters. To quantify the results of Information Retrieval we used evaluation measures like Precision and Recall and also studied how to calculate them effectively. Since the degree of effectiveness greatly depends on the users effort so we discussed how to model the users effort using gain function and discount function of DCG (Discount Cumulative Gain Family).

Effectiveness evaluation is an important aspect of research and design of information retrieval systems. Much research has been done on the topic, and more continues to appear every year. The issue of cost-effective relevance judging and evaluation remains important. Interest in devising user models for evaluations that go beyond individual, independent document relevance has recently increased; ongoing work in novelty and diversity is investigating the tradeoffs between the relevance of documents and the redundancy of relevant information within the documents.

## 7. FUTURE SCOPE
The present Information Retrieval Systems are effective enough to retrieve the relevant pages but still there are some open problems that we discussed like whether these pages are the result of exhaustive search from the Web, how to uniformly sample Web Pages on a Web Site if one does not have complete list of Web Pages.

Also, we know lots of resources are wasted (memory and time) for dealing with duplicate pages so while finding the duplicate pages we also need to work on finding the pages which are semantic duplicates of each other.

## 8. REFERENCES

[1] James M. Abello, Panos M. Pardalos, Mauricio G. C. Resende "Algorithmic Aspects of Information Retrieval on the Web" in "Handbook of Massive Data Sets", Kluwer Academic Publisher, 2002, pp 3-10

[2] Sergey Brin and Lawrence Page "The Anatomy of a Large Scale Hypertextual web Search Engine", in Proceedings of World-Wide Web'98.

[3] J.Klenberg, "Authoritative sources in a hyperlinked environment" Proc. ACM-SIAM Symposium on Discrete Algorithms (1998).

[4] A. Z. Broder (1997). On the resemblance and containment of documents. In Proceedings of Compression and Complexity of Sequences 1997, pp 21-29. IEEE Computer Society.

[5] J.Cho, H.Garcia-Molina, L.Page (1998). Efficient Crawling through URl Ordering. In (WWW7, 1998) pp 161-172

[6] Monika R. Henzinger (2004). Algorithmic Challenges in Web Search Engines. Internet Mathematics Vol. 1, No. 1: 115-126.

[7] K. Bharat and A.Z. Broder (1998). A technique for measuring the relative size and overlap of public web search engines. In (WWW7, 1998) pp 379-388

[8] S. Lawrence and C.L. Giles (1998) Searching the World Wide Web. Science, 280(5360):98.

[9] M. Henzinger, A. Heydon, M. Mitzenmacher, and M. Najork. "On Near-Uniform URL Sampling." In Proceedings of the 9th International World Wide Web Conference, pp. 295—308. Amsterdam: Elsevier Science, 2000.

[10] G.Pandurangan, P. Raghavan, and E. Upfal. "Using Page Rank to Characterize Web Structure." To appear in the 8th Annual International Computing and Combinatorics Conference, 2002.

[11] Cleverdon CW, Mills J (1997) The testing of index language devices. In: Spärck Jones K, Willett P (eds) Readings in information retrieval. Morgan Kaufmann, San Francisco, pp 98–110.

[12] Robertson S, Hull DA (2000) The TREC-9 filtering track final report. In: Voorhees EM, Harman DK (eds) Proceedings of the 9th text retrieval conference (TREC-9) Nov 2000. NIST, Gaithersburg.

[13] Buckley C, Dimmick D, Soboroff I, Voorhees E (2006) Bias and the limits of pooling. In: Dumais ST, Efthimiadis EN, Hawking D, Järvelin K (eds) SIGIR '06: Proceedings of the 29[th] annual international ACM SIGIR conference on research and development in information retrieval, August 6– August 11. ACM, New York, pp 619–620

[14] Lupu M, Piroi F, Huang X, Zhu J, Tait J (2009) Overview of the TREC 2009 chemical IR track. In: Voorhees EM, Buckland LP (eds) Proceedings of the 18th text retrieval conference (TREC 2009), Nov 2009. NIST, Gaithersburg

[15] L Page, S Brin, R Motwani "The Page Rank Citation Ranking: Bringing Order to the Web", Technical Report. Stanford Infolab, 1999, pp 3-4.

[16] Ben Carterette and Ellen M. Voorhees "Overview of Information Retrieval Evaluation" in Current Challenges in Patent Information Retrieval, Chapter 3 2011 Springer, pp 71-76, 81-82.